\documentclass[aps,prd,showpacs,nofootinbib,superscriptaddress,preprintnumbers,twocolumn]{revtex4-1}
\usepackage{graphicx,amsmath,amssymb,soul,bbm,dcolumn,booktabs, multirow,array,float,enumitem,lmodern,dsfont,microtype,nicefrac}
\usepackage[dvipsnames]{xcolor}
\usepackage[utf8]{inputenc}
\newcolumntype{C}{>{$}c<{$}}
\AtBeginDocument{
\heavyrulewidth=.08em
\lightrulewidth=.05em
\cmidrulewidth=.03em
\belowrulesep=.65ex
\belowbottomsep=0pt
\aboverulesep=.4ex
\abovetopsep=0pt
\cmidrulesep=\doublerulesep
\cmidrulekern=.5em
\defaultaddspace=.5em
}
\newcommand{\be}{\begin{eqnarray}}
\newcommand{\ee}{\end{eqnarray}}

\def\fm {\mathop{\hbox{fm}}}
\def\MeV {\mathop{\hbox{MeV}}}

\def\Re {\mathop{\hbox{Re}}}

\def\beq{\begin{equation}}
\def\eeq{\end{equation}}
\def\beqs#1\eeqs{\beq\begin{split} #1 \end{split}\eeq}

\def\comment#1{}

\def\av#1{ \left\langle #1 \right\rangle }

\usepackage{array}   
\newcolumntype{L}{>{$}l<{$}} 
\newcolumntype{S}{>{\footnotesize $}l<{$\normalsize}} 

\def\*#1{\mathbf{#1}}

\usepackage[colorlinks=true,backref=false, linktocpage=true,
citecolor=NavyBlue,urlcolor=NavyBlue,linkcolor=NavyBlue,pdfpagemode=UseOutlines]{hyperref}
\hypersetup{%
  bookmarksnumbered=true,
  pdftitle = {},
  pdfsubject = {},
  pdfauthor = {},
  pdfkeywords = {}
}

\begin{document}
\title{
A cross-channel study of pion scattering from lattice QCD 
}

\author{M.\ Mai}
\email{maximmai@gwu.edu}
\affiliation{The George Washington University, Washington, DC 20052, USA}

\author{C.\ Culver}
\email{chrisculver@email.gwu.edu}
\affiliation{The George Washington University, Washington, DC 20052, USA}

\author{A.\ Alexandru}
\email{aalexan@gwu.edu}
\affiliation{The George Washington University, Washington, DC 20052, USA}
\affiliation{Department of Physics, University of Maryland, College Park, MD 20742, USA}

\author{M.\ D\"oring}
\email{doring@gwu.edu}
\affiliation{The George Washington University, Washington, DC 20052, USA}
\affiliation{Thomas Jefferson National Accelerator Facility, Newport News, VA 23606, USA}

\author{F.\ X.\ Lee}
\email{fxlee@gwu.edu}
\affiliation{The George Washington University, Washington, DC 20052, USA}

\begin{abstract}
We use a chiral model for pion interactions, in the inverse amplitude formalism, to perform a simultaneous analysis of lattice QCD results for pion-pion scattering in all three isospin channels. The input is the finite-volume two-pion spectrum computed using lattice QCD from six ensembles on lattices elongated in one of the spatial dimensions. A two-flavor dynamical lattice QCD action is used with two quark masses corresponding to a pion mass of 315 MeV and 224 MeV. The spectrum in the elastic region is subjected to a global fit which takes into account full correlations across isospin, pion mass and decay constant. The parameters from the fit are used to perform a chiral extrapolation to the physical point. The cross-channel fit results in a more precise determination of the parameters of the model when compared with single channel fits. We obtain, $m_\pi a_0^{I=0}=0.2132(9)$, and $m_\pi a_0^{I=2}=0.0433(2)$ as well as $m_\sigma=443(3)-i221(6)$~MeV and $m_\rho=724(4)-i67(1)$~MeV. Several aspects of scale setting and consistency with previous analyses of lattice QCD results are discussed as well.
\end{abstract}

\pacs
{
12.38.Gc, 
14.40.-n, 
13.75.Lb  
}
\maketitle

\section{Introduction}\label{sec:intro}
Lattice QCD calculations provide an ab-initio access to particle scattering subject to the strong interaction. Calculations of two-hadron states are performed in a small cubic volume with periodic boundary conditions and they are connected analytically to infinite-volume  scattering amplitudes via mapping established in Refs.~\cite{Luscher:1985dn, Luscher:1986pf, Luscher:1990ux}.
In this context, pion-pion scattering has been a prime subject for lattice QCD calculations, in isospin $I=2$~\cite{Sharpe:1992pp, Kuramashi:1993ka, Gupta:1993rn, Yamazaki:2004qb, Aoki:2005uf, Beane:2005rj, Beane:2007xs, Feng:2009ij, Yagi:2011jn, Fu:2013ffa,
Sasaki:2013vxa, Kawai:2017goq, Dudek:2012gj, Bulava:2016mks, Akahoshi:2019klc, Helmes:2015gla, Helmes:2019dpy, Horz:2019rrn}, $I=1$~\cite{Aoki:2007rd, Feng:2010es, Gockeler:2008kc, Lang:2011mn, Bali:2015gji, Guo:2016zos, Pelissier:2012pi, Aoki:2011yj, Dudek:2012xn, Feng:2014gba, Metivet:2014bga, Wilson:2015dqa, Alexandrou:2017mpi, Andersen:2018mau, Fu:2016itp, Alexandrou:2017mpi, Werner:2019hxc}, and $I=0$~\cite{Briceno:2016mjc, Briceno:2017qmb, Guo:2018zss, Fu:2017apw, Liu:2016cba}.

Since most lattice QCD calculations are carried out, for technical reasons, using quark masses heavier than the physical ones,
extrapolation of lattice QCD results to the physical point requires a model for energy and pion mass dependence. Chiral Perturbation Theory (ChPT) allows for the controlled expansion of the $\pi\pi$ scattering amplitude in quark masses and meson momenta.  The ChPT expansion is valid only in the non-resonant region around the threshold. Resonances such as  $\rho(770)$ and $f_0(500)$ (or `$\sigma$'), but also the high energy behavior of the scattering amplitudes, require a non-perturbative treatment, usually guided by imposing constraints from analyticity and two-body unitarity. The inverse amplitude method (IAM)~\cite{Truong:1988zp, Pelaez:2006nj, GomezNicola:2007qj, Pelaez:2010fj} provides a well-established framework in this sense. It is unitary and matches the chiral pion-pion amplitude~\cite{Gasser:1984gg,Gasser:1983yg} up to the next-to-leading order (NLO).

The IAM describes different isospin channels within the same approach; the fit parameters in form of low-energy constants, pion masses, and pion decay constants enter different channels and connect them. Here, for the first time we apply this method to lattice QCD data in different isospin channels determined from the \emph{same} ensembles~\cite{Guo:2018zss, Guo:2016zos, Culver:2019qtx}, thus minimizing the effect of data inconsistencies.

In comparison to previous studies in which the IAM was applied~\cite{Hu:2017wli, Doring:2016bdr, Hu:2016shf, Bolton:2015psa} we allow here the pion mass and decay constant to vary and include their full correlations with the two-hadron energy levels (ELs) in the fit; for the isospin $I=2$ channel alone, this was recently achieved in Ref.~\cite{Culver:2019qtx}. In addition, the correlations of ELs between {\it different} isospin channels are taken into account, for the first time. We work in lattice units throughout the paper up to the last step, when evaluating the amplitude at the physical pion mass, which involves a scale setting. 

The global fit allows one to address issues that remained unresolved in previous studies. For example, using SU(3) unitarized ChPT with contact terms~\cite{Oller:1998hw}, which can only be approximately matched to NLO SU(3) CHPT, it was found that data from $N_f=2$ lattice QCD calculations by different groups~\cite{Bali:2015gji, Guo:2016zos, Gockeler:2008kc, Lang:2011mn} extrapolate to consistently low physical $\rho$ masses~\cite{Hu:2016shf} (see Refs.~\cite{GomezNicola:2001as, Nebreda:2010wv, Guo:2012ym, Guo:2015xva} for more recent one-loop SU(3) versions). Within the model, the $K\bar K$ channel could explain the discrepancy, but also missing higher orders in the model could be responsible, or systematic effects on the side of lattice QCD calculations:  Small volumes could induce large exponential corrections that cannot be systematically represented in models for the scattering region.  Scale setting can be another source of discrepancy in lattice QCD data. See also Ref.~\cite{Hu:2017wli} for a corresponding analysis of $N_f=2+1$ lattice QCD calculations. 
Here, we have the opportunity to analyze different isospin channels at the same time with a model that matches to NLO two-flavor ChPT. This will allow us to avoid some of the potential pitfalls at least for the results computed by the GW lattice QCD group~\cite{Guo:2018zss, Guo:2016zos, Culver:2019qtx}.

\section{Lattice QCD details }\label{sec:lattice}

The input used in this study are two-hadron state energies in a finite box with periodic boundary conditions computed using lattice QCD. Our results use two different quark masses corresponding to a pion mass of $m_\pi=224\MeV$ and $m_\pi=315\MeV$. For each quark mass we use three different box geometries to scan the energy region below the inelastic threshold in each channel. The parameters for each of the six lattice ensembles are listed in Table~\ref{table:gwu_lattice}. 

All lattice results are computed using QCD with two mass-degenerate quark flavors ($N_f=2$ QCD). This is a good approximation of the real world when focusing on the lightest hadrons which are composed mainly of up and down quarks, as is the case in this study. Moreover, $N_f=2$ QCD is also a very interesting theoretical model with a minimal number of parameters: one quark mass and $\Lambda_\text{QCD}$. This can be used as a precise testbed for non-perturbative aspects of QCD. To generate the ensembles, both gauge and quark actions use improved discretizations. For the gauge action we use L\"uscher-Wise action~\cite{Luscher:1984xn} and for the quarks we use nHYP discretization~\cite{Hasenfratz:2007rf}.

The data analysed here correspond to two-pion states for all three possible isospin combinations. Specific details can be found in Refs.~\cite{Pelissier:2012pi,Guo:2018zss} for $I=1$, Ref.~\cite{Guo:2016zos} for $I=0$, and Ref.~\cite{Culver:2019qtx} for $I=2$. Here we review briefly the lattice methods used to compute the two-pion state energies and the other relevant observables used in this study.

\subsection{Two-pion finite volume spectrum}\label{sec:fields}

The spectrum of two-hadron states in a box with periodic boundary conditions is quantized. The energy levels, in the elastic region, can be related with the scattering amplitude in the infinite volume. To extract the energy levels we use the standard {\em variational} method~\cite{Luscher:1990ck}.
For each isospin channel we construct a set of interpolating fields that are expected to have large overlap with the lowest lying states in the spectrum. We include sufficient interpolators to resolve all the states with energies below the inelastic threshold. The interpolators $\mathcal{O}_i$ are used to construct a correlator matrix, 
\beq
    C_{ij}(t)=\av{\mathcal{O}_i(t)\mathcal{O}_j^{\dagger}(0)}\,.
\eeq
The eigenvalues are extracted by solving the generalized eigenvalue problem,
\beq
    C(t_0)^{-\frac{1}{2}}C(t)C(t_0)^{-\frac{1}{2}}\psi^{(n)}(t,t_0)=\lambda^{(n)}(t,t_0)\psi^{(n)}(t,t_0)\,.
\eeq
Here $t_0$ is a parameter that is adjusted for each isospin to help dampen the effects of excited state contributions.  It was shown in Refs.~\cite{Luscher:1990ck, Blossier:2009kd}, that the energies of the system can be extracted from the long-time behavior of the generalized eigenvalues 
\beq
\lambda^{(n)}(t,t_0)\propto e^{-E_nt}\left[1 + \mathcal{O}(e^{-\Delta E_nt})\right]\,.
\eeq

Finite volume states will inherit the symmetries of the box and they can be classified according to the irreducible representations (irreps) of the box symmetry group. To properly identify the states corresponding to the energy levels extracted from the variational analysis, we need to design interpolating fields that have the appropriate transformation properties. For this study we use both cubic boxes and boxes that are elongated in one of the dimensions. The relevant lattice symmetry group for the cubic box is $O_h$ whose 10 irreps are conventionally named as $A^\pm_1, A^\pm_2, E^\pm, T^\pm_1, T^\pm_2$, and for the elongated box $D_{4h}$ whose 10 irreps are $A^\pm_1, A^\pm_2, E^\pm, B^\pm_1, B^\pm_2$.  In the energy range we study, scattering in the $I=0$ and $I=2$ is dominated by the $\ell=0$ partial wave, and for $I=1$ the $\ell=1$ partial wave. In a finite box, different partial waves are mixed by the finite volume effects.  The rotationally-symmetric SO(3) angular momentum multiplets in the continuum are split into multiplets transforming under the symmetry group of the box.  The splitting is shown in Table~\ref{table:irep_split} (for details see Ref.~\cite{Lee:2017igf}).
We see that the relevant lattice irreps for $I=0,2$ channels are $A_1^+$ in both box symmetries. For the $I=1$ channel we use the $A_2^-$ irrep which is sensitive to the elongation.

\begin{table*}[t]                   
\begin{ruledtabular}
\begin{tabular}{@{}*{13}{>{$}l<{$}}@{}}                                                                
\text{ensemble}~& ~N_t\times N_{x,y}^2\times N_z~ & ~\eta~ & ~a[\fm]~   & ~N_\text{cfg}~  & ~am_{\pi}~  & ~am^{pcac}_{u/d}~  & ~af_{\pi}~  \\
\midrule                                                                                               
\mathcal{E}_1&48\times24^2\times24  &  1.00  & 0.1210(2)(24) & 300   & 0.1931(4)  & 0.01226(5) & 0.0648(8)~    \\ 
\mathcal{E}_2&48\times24^2\times30  &  1.25 &         & 300     &    0.1944(3)      &     0.01239(4)     &  0.0651(6)~      \\
\mathcal{E}_3&48\times24^2\times48  &  2.00  &         & 300    &    0.1932(3)   &     0.01227(5)     &  0.0663(6)~         \\
\mathcal{E}_4&64\times24^2\times24  &  1.00  & 0.1215(3)(24)& 400  & 0.1378(6)  & 0.00612(5)~& 
0.0600(10)  \\                                                                                           
\mathcal{E}_5&64\times24^2\times28  &  1.17 &         & 378     & 0.1374(5) &   0.00620(4)       &   0.0600(8)~         \\
\mathcal{E}_6&64\times24^2\times32 &  1.33 &         & 400     & 0.1380(5)      &  0.00619(4)      &   0.0599(10)      \\
\end{tabular}  
\end{ruledtabular}
\caption{Summary of the ensembles used in all isospin channels.  Included for reference are the lattice geometry and elongation~$\eta$, the lattice spacing~$a$, the number of configurations in each ensemble, the pion mass, the current quark mass (see Eq.~\ref{eq:mpcac}), and the pion decay constant.}
\label{table:gwu_lattice}                                                                                  
\end{table*}    

\begin{table}[t]
\begin{tabular}{r c c}
\toprule
$~~\ell~~~~~$ & $O_{h}$ & $D_{4h}$ \\
\hline
~~0~~~~~ & $A_1^+$                                        & $A_1^+$\\
~~1~~~~~ & $T_1^-$                                        & $A_2^-\oplus E^-$\\
~~2~~~~~ & $E^+ \oplus T_2^+$                             & $A_1^+\oplus B_1^+ \oplus B_2^+ \oplus E^+$\\
~~3~~~~~ & $A_2^- \oplus T_1^- \oplus T_2^-$              & $A_2^- \oplus B_1^- \oplus B_2^- \oplus 2 E^-$\\
~~4~~~~~ & $A_1^+ \oplus E^+ \oplus T_1^+ \oplus T_2^+$ & $2A_1^+ \oplus A_2^+ \oplus B_1^+ \oplus B_2^+ \oplus 2E^+$\\
\bottomrule
\end{tabular}
\caption{Resolution of angular momentum in terms of irreps of the $O_h$ and the $D_{4h}$ group.}
\label{table:irep_split}
\end{table}

Once the appropriate operators are determined, correlation functions are computed using Wick contractions on the quark fields.  The resulting quark-diagrams depend on the isospin channel -- the details are included in the references listed above. The correlation functions require the evaluation of the all-to-all quark propagator, that is the quark propagator from every point on the lattice to any other point.  To avoid the full expense of this calculation we use the Laplacian-Heaviside method (LapH)~\cite{Peardon:2009gh}.  The idea is to truncate the quark interpolating fields by dropping out the high-frequency modes of the three-dimensional Laplacian on each time slice while preserving the symmetry of the resulting ``smeared'' fields.  The interpolating fields constructed out of these quarks excite the same QCD states, but overlap better with the low-energy states. An additional advantage is that we only need to invert the Dirac matrix for the LapH modes, reducing the numerical cost.  For all isospin channels we used the $N_v=100$ lowest eigenvectors corresponding to a smearing radius of approximately $0.5\fm$~\cite{Guo:2016zos}.  The smeared quark propagators were computed efficiently using a set of GPU inverters~\cite{Alexandru:2011ee}.  

\subsection{Other observables}

Besides the energy of the two-pion states, the other lattice QCD inputs for the analysis are the pion mass and the pion decay constants. These parameters for each lattice ensemble are listed in Table~\ref{table:gwu_lattice}.  We note that all ensembles are generated with the same coupling, which should generate the same lattice spacing (or cutoff).  For ensemble $\mathcal{E}_{1,2,3}$ the quark mass is the same and similarly for ensembles $\mathcal{E}_{4,5,6}$. The differences in pion mass within these sets is thus expected to be the result of statistical fluctuation and/or finite volume effects.  The pion mass was computed by evaluating the two-point function of the pion using LapH with a $\bar{u}\gamma_5 d$ interpolating field. 

The value of the pion decay constant $f_{\pi}$ was computed using standard methods (see for example Ref.~\cite{Brandt:2013dua}). We use two two-point correlation functions: $\av{A_4(t)P(0)^\dagger}$ and $\av{P(t)P(0)^\dagger}$, where $P$ is the pseudo-scalar density $\bar{q}\gamma_5 q$ and $A_4$ is the axial current density $\bar{q}\gamma_4\gamma_5 q$. Both $P(t)$ and $A_4(t)$ are projected to zero spatial momentum. From the $\av{P(t)P(0)^\dagger}$ correlation function the pion mass $m_{\pi}$ and overlap factor $Z$ are extracted:
\begin{equation}
\av{P(t)P(0)^\dagger} \xrightarrow{t\to\infty} \frac{Z^2}{2m_\pi} e^{-m_\pi t}\,.
\end{equation}
The ratio of the two correlation functions is used to calculate the current quark mass 
\beq
\label{eq:mpcac}
m_\text{PCAC} \equiv \frac12 \frac{\av{\partial_t A_4(t) P(0)^\dagger}}
{\av{P(t)P(0)^\dagger}}\,,
\eeq
where $\partial_t A_4(t)\equiv[A_4(t+1)-A_4(t-1)]/2$.
Using these values the pion decay constant is defined as 
\begin{equation}
f_{\pi}\equiv\sqrt{2 Z^2}\,\frac{m_{\text{PCAC}}}{m_{\pi}^2}\,.
\end{equation}
The decay constant needs to be renormalized, but the renormalization is expected to introduce only a couple of percent shift~\cite{Hoffmann:2007nm}.

\begin{figure*}[p]
    \centering
    \begin{tabular}{c|c|c}
    \multicolumn{3}{c}{Finite-volume spectrum}\\
    \toprule
    \multicolumn{1}{c}{$I=0,~A_1^+$} &\multicolumn{1}{c}{$I=1,~A_2^-$} &\multicolumn{1}{c}{$I=2,~A_1^+$}\\
    \multicolumn{1}{c}{$\mathbf{P=0}$~~~~~~~~~~~~~~~~~~$\mathbf{P=1}$} &\multicolumn{1}{c}{$\mathbf{P=0}$~~~~~~~~~~~~~~~~~$\mathbf{P=1}$}  &\multicolumn{1}{c}{~~~$\mathbf{P=0}$~~~~~~~~~~~~~~~~~$\mathbf{P=1}$} \\[-0.2cm]
     \includegraphics[width=0.30\linewidth,trim=0 2.9cm 28.2cm .5cm,clip]{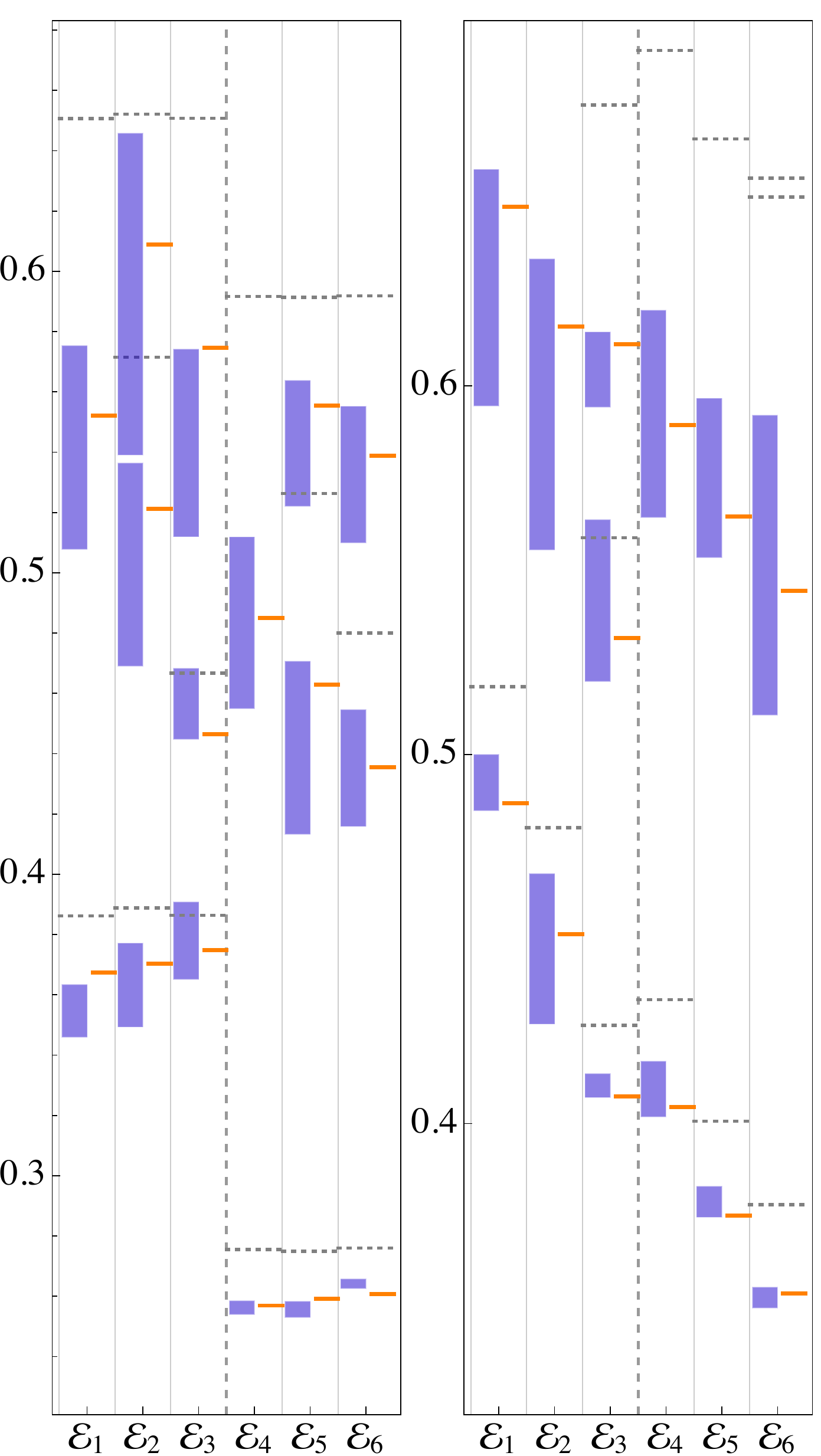}~~~
    &\includegraphics[width=0.30\linewidth,trim=14.1cm 2.9cm 14.1cm .5cm,clip]{DNA-final.pdf}~~~
    &\includegraphics[width=0.30\linewidth,trim=28.3cm 2.9cm 0 .5cm,clip]{DNA-final.pdf}\\
    \multicolumn{3}{c}{}\\
    \multicolumn{3}{c}{Infinite-volume spectrum}\\
    \toprule
    \multicolumn{1}{c}{$I=0,~l=0$} &\multicolumn{1}{c}{$I=1,~l=1$} &\multicolumn{1}{c}{$I=2,~l=0$}\\[-0.4cm]
    \parbox[c]{6.2cm}
        {
        \includegraphics[width=\linewidth, trim=0 1.75cm 0 0,clip]{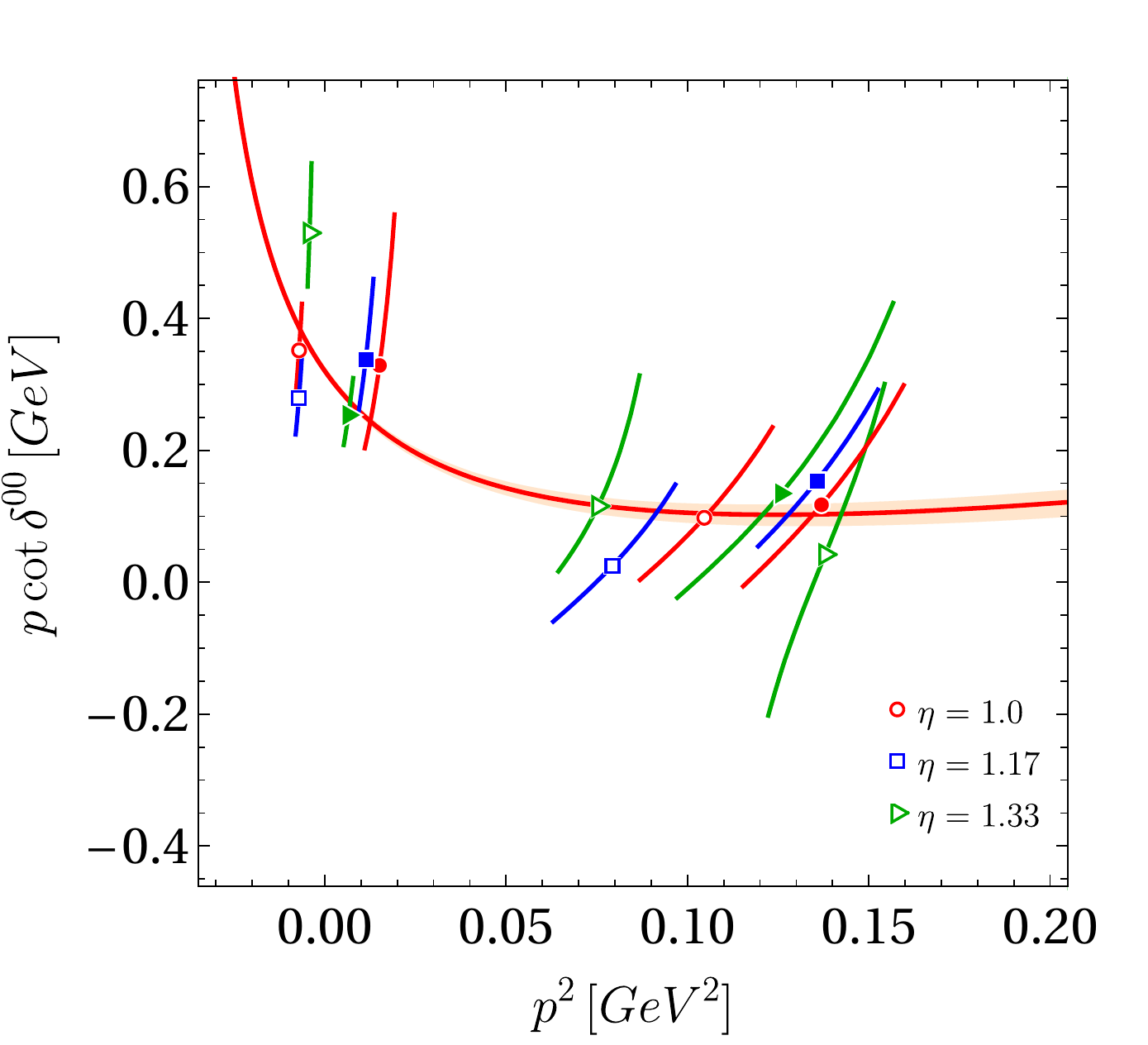}\\
        \includegraphics[width=\linewidth, trim=0 0      0 0.7cm,clip]{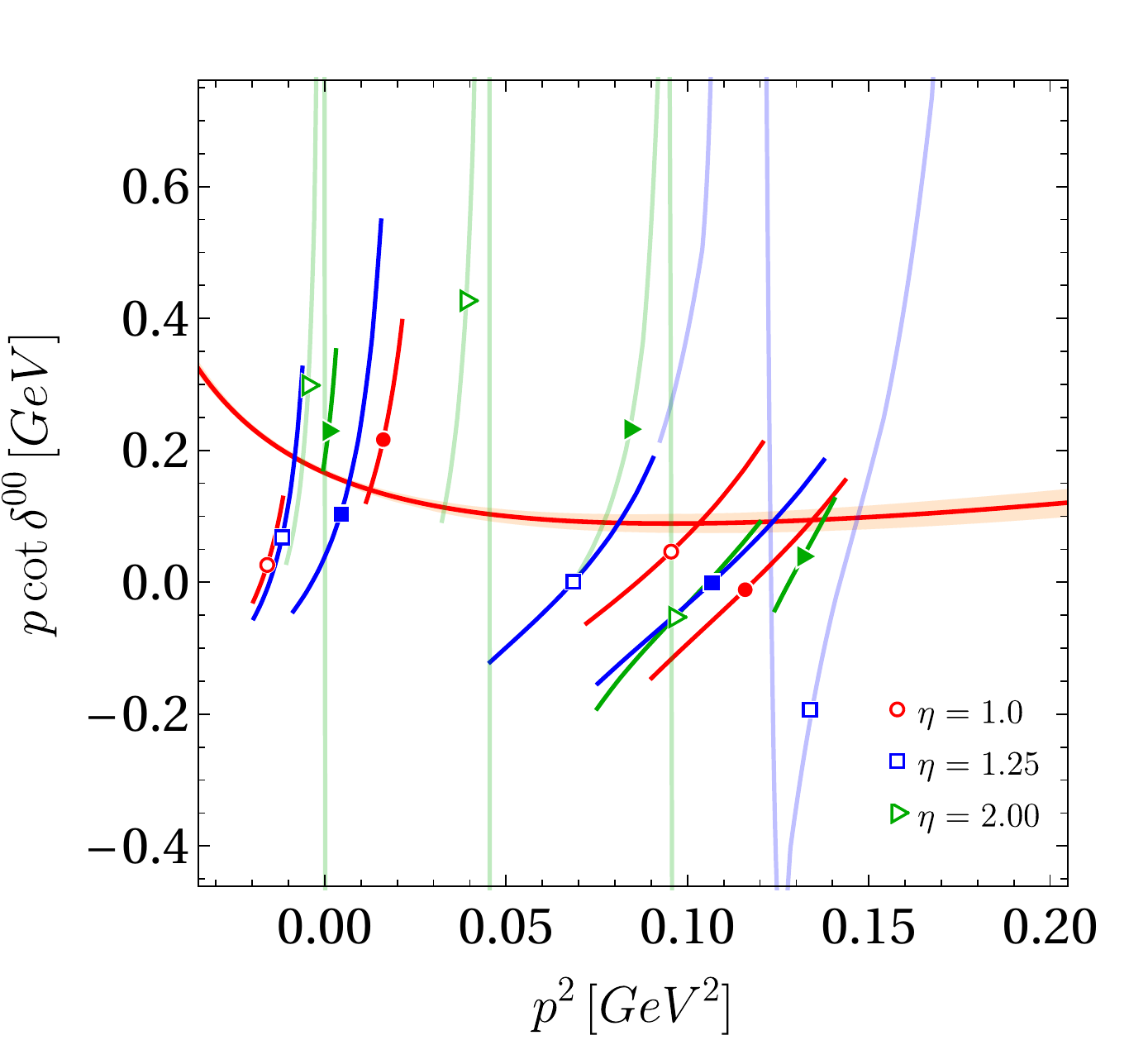}
        }
    &
    \parbox[c]{6.0cm}{
        \includegraphics[width=\linewidth, trim=0 1.75cm 0 0,clip]{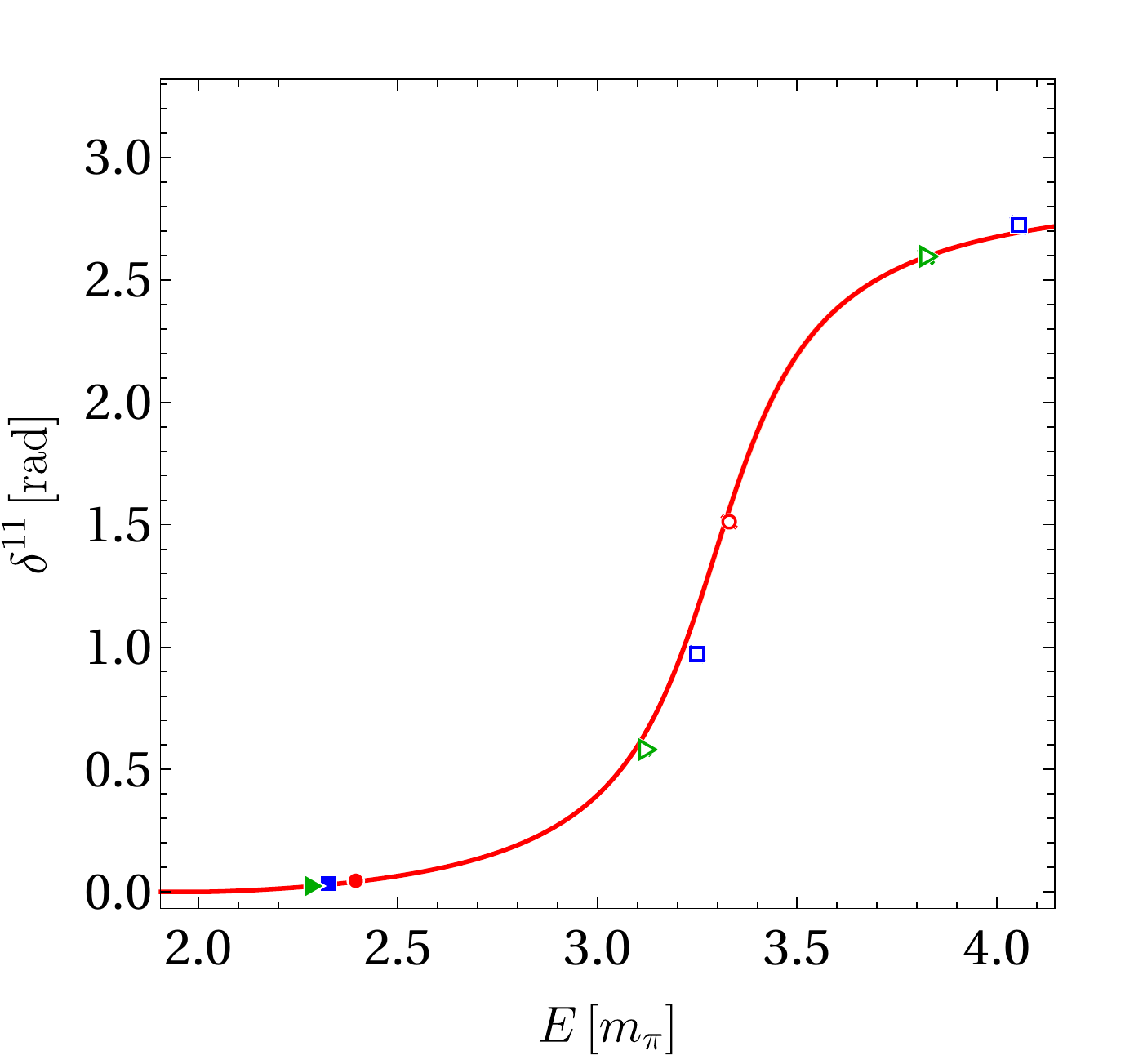}\\
        \includegraphics[width=\linewidth, trim=0 0      0 0.7cm,clip]{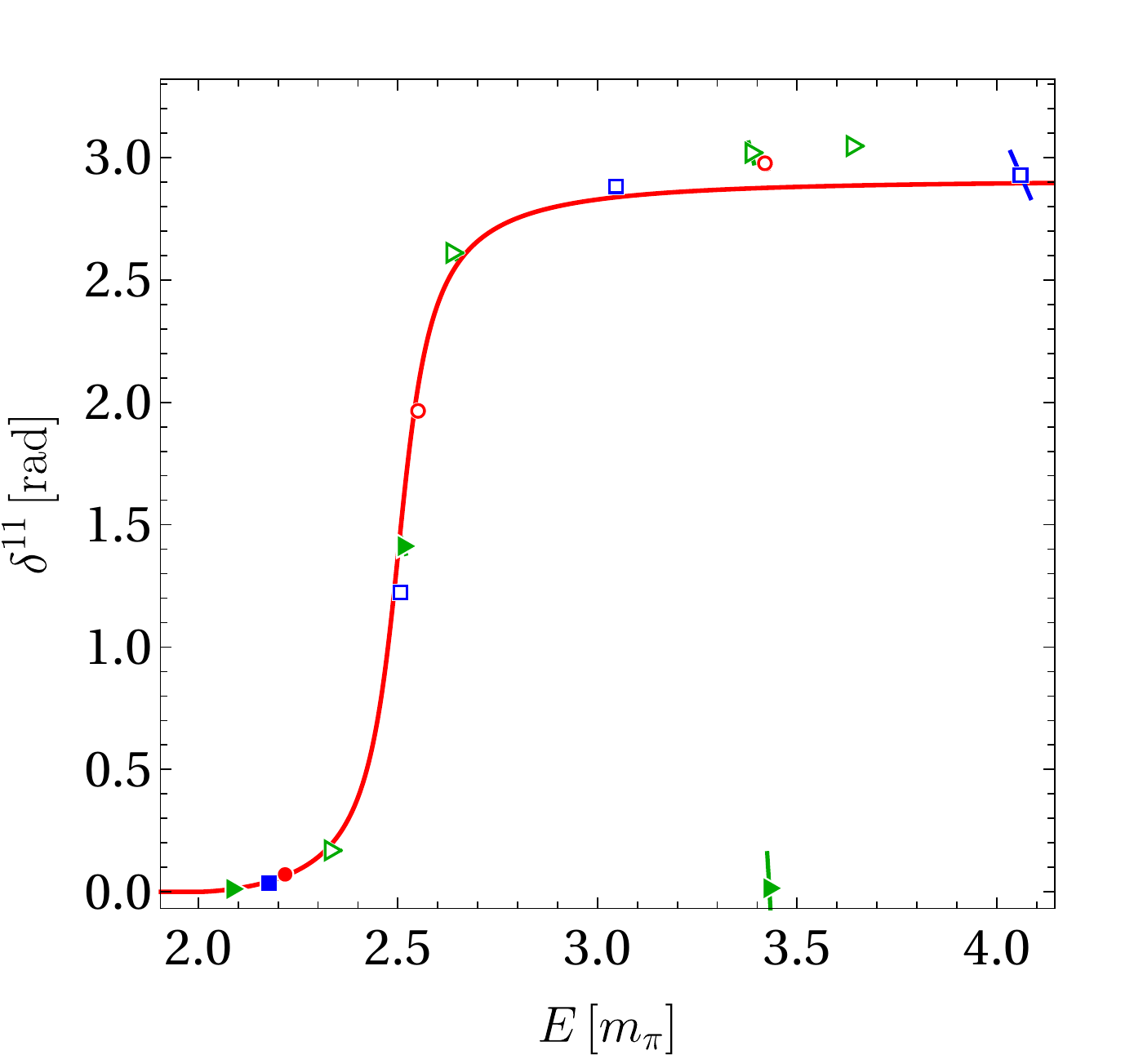}}
    &
    \parbox[c]{6.2cm}{
        \includegraphics[width=\linewidth, trim=0 1.75cm 0 0,clip]{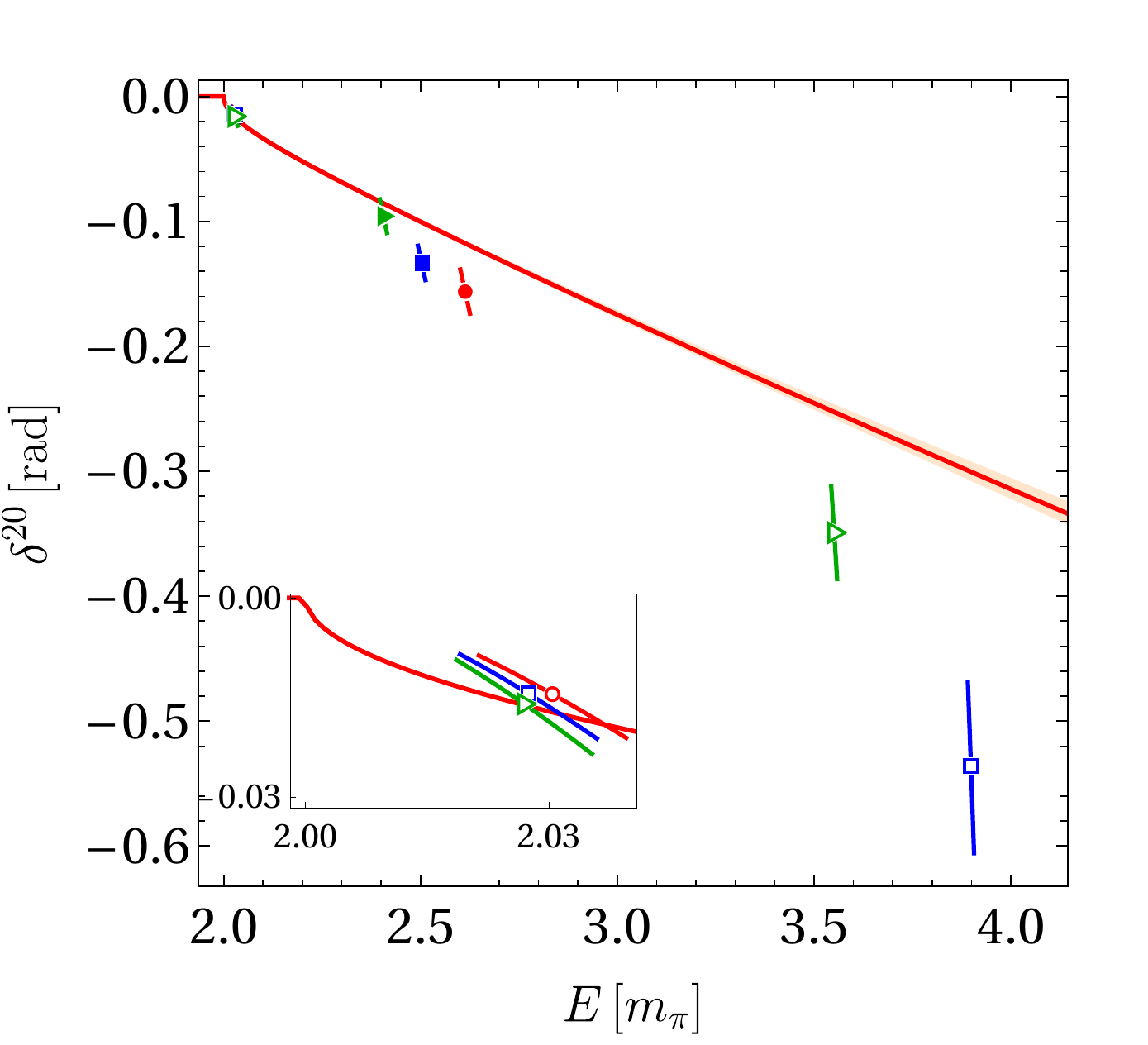}\\
        \includegraphics[width=\linewidth, trim=0 0      0 0.7cm,clip]{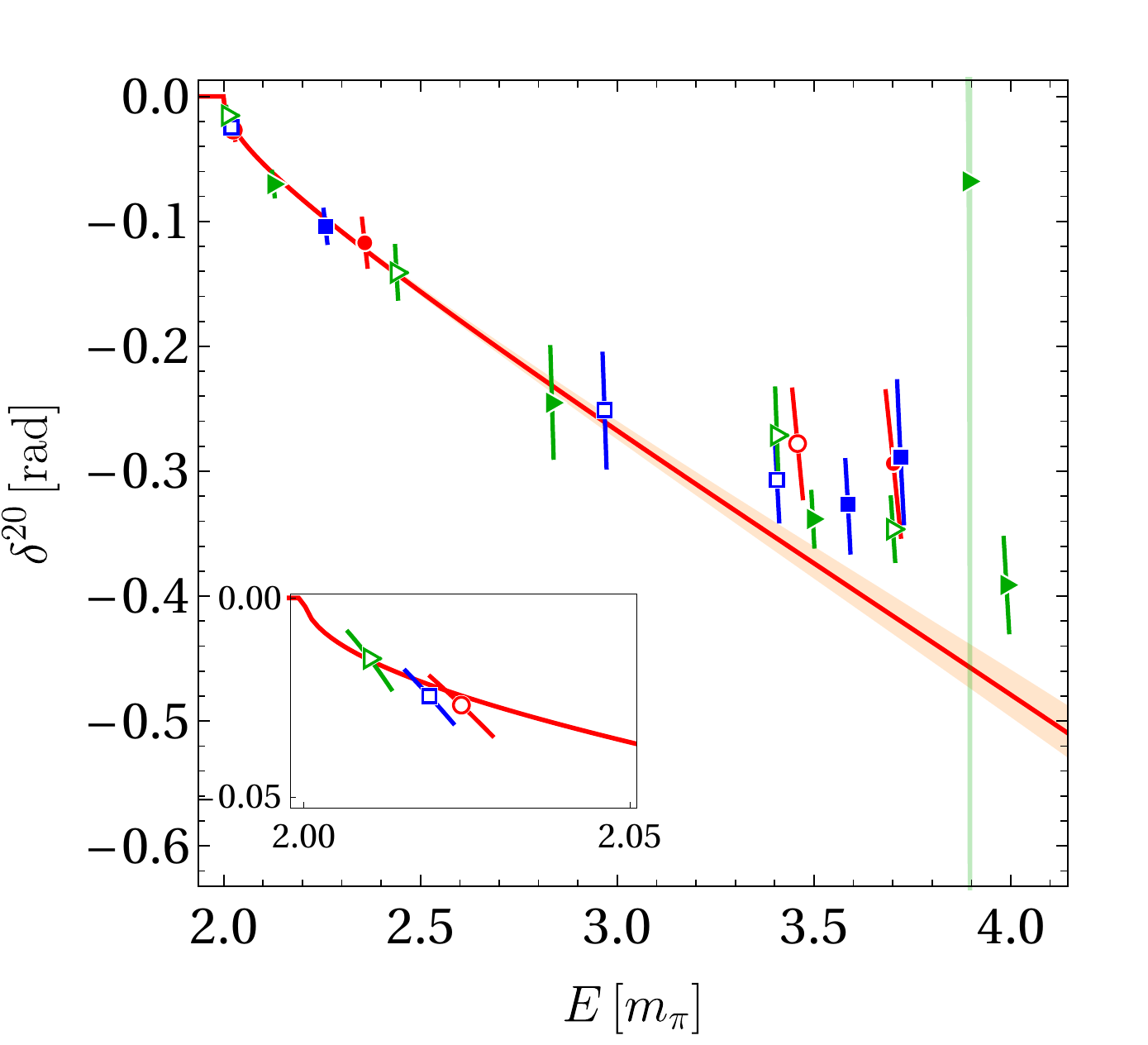}}
\end{tabular}
\caption{
\label{fig:DNA}
\underline{Top panel:} Energy eigenvalues (light blue bars) of all $\pi\pi$ channels determined on 6 gauge configurations - $\{\mathcal{E}_1,..\mathcal{E}_6\}$ for a given isospin, irrep and boost momentum $P$. The horizontal gray and orange lines denote the location of non-interacting levels and that of the central value of the global fit to these data. \underline{Lower panel}: Phase-shifts in all three $\pi\pi$ channels after mapping finite-volume spectrum (upper panel) and the global fit results using L\"uscher's method for a given isospin and angular momentum. The orange bands show the uncertainty of the global fit.
}
\end{figure*}

For each pion mass we use one cubic box, and two elongated boxes. The elongated boxes are important for getting a good scan of the relevant scattering region in each isospin channel. The lowest momentum of a particle moving in the elongated direction is $\frac{2\pi}{L\eta}$.  The energy of the particle thus changes as we vary $\eta$.  For two-pion states with total momentum $\*P=[000]$ the energy changes little with $\eta$ unless the state corresponds to two pions with non-zero relative momentum, which tends to be the case for the excited levels in each channel.  For states with total momentum $\*P=[100]$ in the elongated direction, even for the lowest state the energy usually varies with $\eta$ and the energy levels cover well the kinematic region of interest.  This is indeed observed in the extracted spectrum shown in Figure~\ref{fig:DNA}.

\section{Global study of \texorpdfstring{$\pi\pi$}{} scattering}\label{sec:analysis}

The determination of the $I=2$ finite volume spectrum from lattice QCD calculations concludes a multi-year program~\cite{Guo:2018zss,Guo:2016zos,Culver:2019qtx} of the GW lattice group in obtaining comprehensive information on $\pi\pi$ scattering from first principles. The obtained set of energy eigenvalues covers a large energy region from below the production threshold to and beyond the resonance region. At the same time these results describe $\pi\pi$ scattering at two unphysical pion masses ($\sim1.5$ and  $\sim2.5~m_\pi^{\rm phys}$). Thus, they provide a unique opportunity for mapping out the $m_\pi$ vs. $E$ plane with respect to $\pi\pi$ interactions in all three isospins, which is explored in the following.

To make full use of the available information, a scattering amplitude is required which not only takes into account the analytic properties in $E$ but also the chiral behavior, consistent with constraints from perturbative ChPT at NLO as well as those from Ref.~\cite{Bruns:2017gix}. A method reconciling both demands is the so-called modified inverse amplitude method (mIAM)~\cite{Truong:1988zp, Dobado:1996ps, GomezNicola:2007qj, GomezNicola:2007qj}. In the past, it has been shown to be very successful in describing experimental data on $\pi\pi$ scattering~\cite{Hanhart:2008mx,Pelaez:2015qba,Nebreda:2010wv} in all three isospin channels, while also having correct chiral behavior up to the next-to-leading chiral (NLO) order. Likewise it fulfills the general requirements on the chiral trajectory for resonances, derived in Ref.~\cite{Bruns:2017gix} to all chiral orders.

\begin{figure}[t!]
    \centering
    \includegraphics[width=0.99\linewidth]{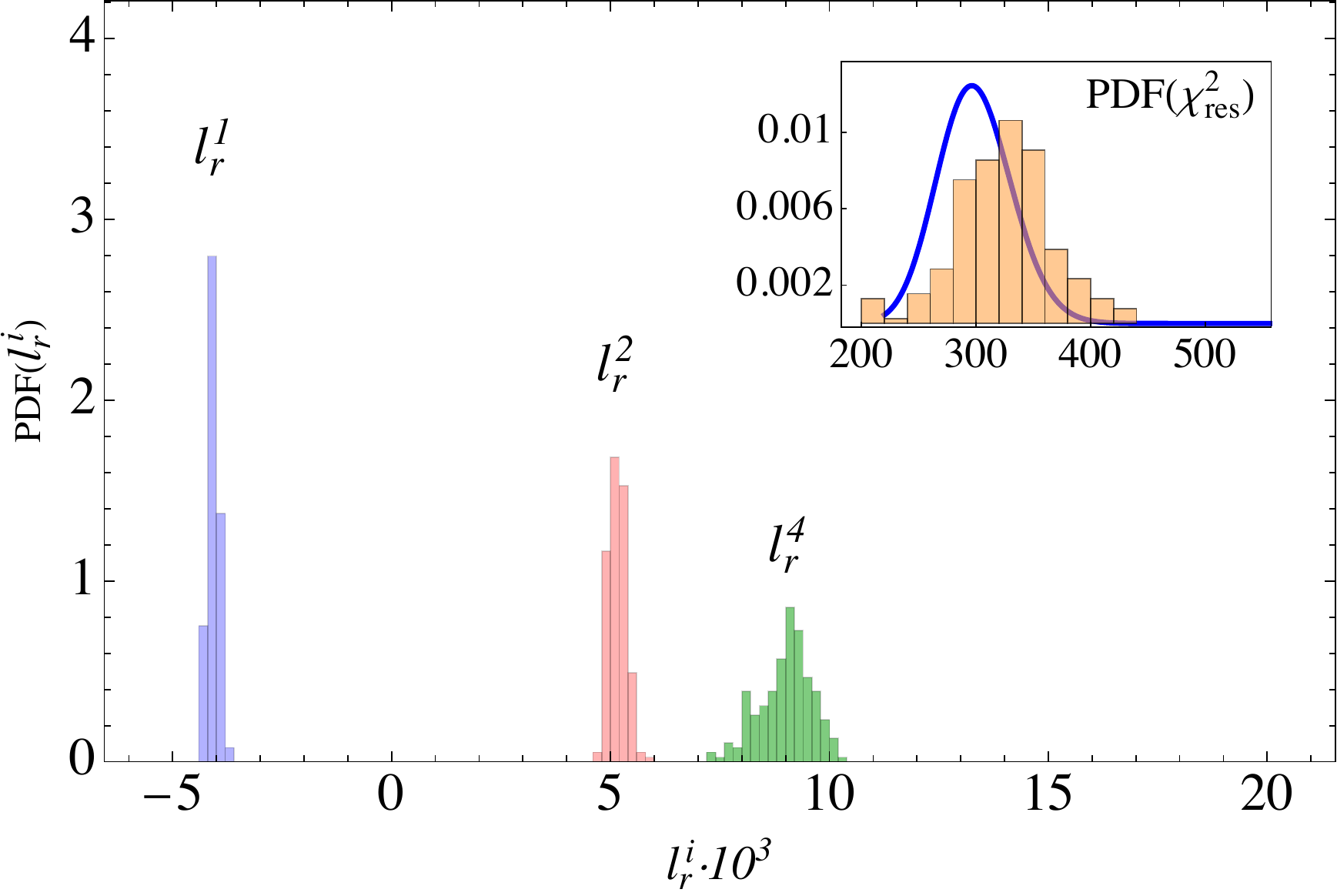}
    \caption{Probability distribution of the LECs as determined in a re-sampling procedure when using mIAM~\cite{Truong:1988zp,FernandezFraile:2007fv,Hanhart:2008mx,Pelaez:2015qba} in a global fit to all scattering channels at both unphysical pion masses. The inset shows the distribution of the the re-sampled $\chi_{\rm res}^2$'s (histogram) together with the prediction from a non-central $\chi^2$ distribution (blue line).}
    \label{fig:ldistribution}
\end{figure}

\subsection{Infinite volume spectrum}\label{sec:formalism}

The modified inverse amplitude method is based on the leading ($T_2^{Il}(s)$) and NLO ($T_4^{Il}(s)$) chiral amplitudes projected to a specific isospin ($I$) and angular momentum ($l$). A unitary scattering amplitude $T_{mIAM}^{Il}(s)$ can then be derived~\cite{Truong:1988zp} using dispersion relations, namely
\begin{align}
\label{eq:mIAM}
 T_{mIAM}^{Il}(s)=\frac{(T_2^{Il}(s))^2}{T_2^{Il}(s)-T_4^{Il}(s)+A^{Il}_m(s)}\,.
\end{align}
The term $A^{Il}_m(s)$ in the denominator does not arise for dynamical reasons, but has been introduced~\cite{FernandezFraile:2007fv,GomezNicola:2007qj} to avoid the appearance of an unphysical pole. Such a pole is associated with appearance of the so-called Adler zero -- a sub-threshold energy at which the amplitude vanishes as demanded by chiral symmetry. Explicitly it reads
\begin{align}
A^{Il}_m(s)&=T^{Il}_4(s_2)\\\nonumber
&~~~~~~~~~-\frac{(s_2-s_A)(s-s_2)}{s-s_A}\left(T^{Il}_2(s_2)-T^{Il}_4(s_2)\right)'\,,
\end{align}
where $s_A$ and $s_2$ are the zeros of $T_2(s)-T_4(s)$ and $T_2(s)$, respectively. With this the $T_{mIAM}$ is an analytic, unitary scattering amplitude, which indeed reproduces the usual chiral expansion and is crossing symmetric up to the next-to-leading chiral order. 

The leading order chiral amplitude is a function of energy, Goldstone-boson mass, $m^2=B(m_u+m_d)$ and pion decay constant in the chiral limit, $f_0$. The amplitude $T_4^{Il}$ involves in the two-flavor case two low-energy constants (LECs) $\bar l_1$ and $\bar l_2$. Two additional low-energy constants $\bar l_3$, $\bar l_4$ enter the NLO chiral amplitude when replacing the above mass and decay constants by their physical values  using one-loop results~\cite{Gasser:1983yg},
\begin{align*}
m_\pi^2=m^2\left(1-\frac{m^2}{32\pi^2f_0^2}\bar l_3\right)
\, ,\,\,
f_\pi=f_0\left(1+\frac{m^2}{16\pi^2f_0^2}\bar l_4\right)\,.
\end{align*}
The constants $\bar l_i$ do not depend on the regularization scale, but only on the parameters of the underlying theory - the quark masses. However, they are related to the scale-dependent, but quark-mass independent renormalized LECs via 
\begin{align}
l_i^r=\frac{\gamma_i}{32\pi^2}\left(\bar l_i+\log \frac{m^2}{\mu^2}\right)\,
\end{align}
where $\gamma_1=\nicefrac{1}{3}$, $\gamma_2=\nicefrac{2}{3}$, $\gamma_3=-\nicefrac{1}{2}$, $\gamma_4=2$. Hence, for a fixed scale $\mu$ one can determine the renormalized LECs and then make predictions for two-particle scattering at a different pion mass. In the course of this work we use dimensional regularization with $\mu=770$~MeV, but emphasize that the amplitude~\eqref{eq:mIAM} is manifestly scale independent.

\begin{table}[t!]
    \centering
    \begin{ruledtabular}
   \begin{tabular}{c|c}
    \addlinespace[0.2em]
    \multicolumn{2}{c}{
    ~~~~$l^r_1=-4.07^{+0.12}_{-0.13}$~~~~
        $l^r_2=+5.14^{+0.23}_{-0.19}$~~~~
        $l^r_4=+9.05^{+0.54}_{-0.70}$~~~~
    }
    \\[0.4em]
    \hline
    \addlinespace[0.4em]
    ~~~ $m_\pi^{\rm light}=223.98^{+0.07}_{-0.13}$~MeV&
        $m_\pi^{\rm heavy}=315.40^{+0.09}_{-0.05}$~MeV~~~\\[+0.2cm]
    ~~~ $f_\pi^{\rm light}= 98.29^{+0.22}_{-0.12}$~MeV&  
        $f_\pi^{\rm heavy}=107.48^{+0.08}_{-0.12}$~MeV~~~\\[0.2em]
    \end{tabular}
    \end{ruledtabular}
    \caption{The fitted LECs ($l^r_i\cdot 10^3$) and $m_\pi$ and $f_\pi$ from the mIAM fit ($\chi^2_{\rm dof}=218/(88-7)$) to all lattice QCD results~\cite{Culver:2019qtx,Guo:2018zss,Guo:2018zss}.}
    \label{tab:lec}
\end{table}

\subsection{Finite volume spectrum}\label{sec:finite volume}

The scattering amplitude introduced in the previous section describes the scattering of two pions in infinite volume in terms of a complex-valued function of continuous energy/momentum variables. In finite volume, momenta are discretized leading to a discretized interaction spectrum. The way to convert the latter into phase-shifts is given by L\"uscher's method~\cite{Luscher:1986pf,Luscher:1985dn}, see also Refs.~\cite{Lee:2017igf,Briceno:2017max}. In the context of mIAM its implementation has been used and is described in Refs.~\cite{Mai:2018djl,Mai:2018xwa}. 

In infinite volume, the scattering amplitude~\eqref{eq:mIAM} is related to the phase-shifts via 
\beq
T_{\rm mIAM}^{Il}(s)=\frac{\sqrt{s}}{2p(\cot\,\delta_{\rm mIAM}^{Il}(s)-i)}\,.
\eeq
We use Eq.~\eqref{eq:mIAM} to compute the phase-shifts predicted by mIAM:
\begin{align}
\cot\,\delta_{\rm mIAM}^{Il}(s)&=\\\nonumber
\frac{\sqrt{s}}{2p}&
\Bigg(\frac{T_2^{Il}(s)-\bar T_4^{Il}(s)+A_m^{Il}(s)}{(T_2^{Il}(s))^2}
-16\pi\Re{J(s)}\Bigg)\,,
\end{align}
where $\bar T_4^{Il}$ denotes the NLO chiral amplitude without $s$-channel loop diagrams and $J(s)$ denotes the meson-meson loop in dimensional regularization for $\mu=770$~MeV. The determination of the corresponding finite-volume spectrum amounts to finding the roots of the following set of equations
\begin{align}  
\label{eq:mIAM-FVspectrum}
p\,
\cot\,\delta^{00}(s)&=
\frac{2\pi}{L}\frac{\mathcal{Z}_{00}(1,q^2;\eta)}{\pi^{3/2}\eta}\,,\\
\cot\,\delta^{11}(s)&=
\frac{\mathcal{Z}_{00}(1,q^2;\eta)}{\pi^{3/2}\eta q}
+\frac{2}{\sqrt{5}}
\frac{\mathcal{Z}_{20}(1,q^2;\eta)}{\pi^{3/2}\eta q^3}\,,\nonumber
\\  
\cot\,\delta^{20}(s)&=
\frac{\mathcal{Z}_{00}(1,q^2;\eta)}{\pi^{3/2}\eta q}\,\nonumber
\end{align}
for the irreps $A_1^+$, $A_2^-$, and $A_1^+$, respectively. In every interaction channel ($I=0,1,2$ from top to bottom, respectively) the right-hand side carries the information on the geometry of the lattice (size $L$, elongation $\eta$), and kinematics via $q=L/(2\pi)p(s)$, where $p$ is the magnitude of the pion relative three-momentum in the center of the mass system. The required L\"uscher functions for elongated boxes as well as corresponding formulas for moving frames are quoted in Ref.~\cite{Lee:2017igf}. The left hand side of the above equations contains only the information on the pion interaction in the corresponding channel.

We note that Eqs.~(\ref{eq:mIAM-FVspectrum}) are valid when neglecting higher partial waves and only below the inelastic threshold (4$m_\pi$). The cutoff in the angular momentum space is justified by the smallness of higher partial waves in all channels. Additionally, note that the factor $p$ on both sides of the first equation makes it well-defined below $\pi\pi$ threshold, where the lattice QCD result also exist, see also previous studies~\cite{Doring:2016bdr,Guo:2018zss,Briceno:2016mjc} of this channel.

\subsection{Fit to finite volume spectrum}\label{sec:fit to finite volume}

As discussed above there are four parameters in the model -- the LECs $l_i^r$. Multiple work-flows are possible when confronting lattice QCD results with the model, i.e. with respect of how many interaction channels are included, or which of the experimental or lattice data are included. After evaluating several such options, we decided to fit to lattice QCD results only and predict $\pi\pi$ scattering at the physical point. This is possible due to the correct chiral properties of the mIAM also in the vicinity of resonances~\cite{Bruns:2017gix}. Additionally, the fitted lattice QCD eigenvalues include correlations\footnote{The covariance matrices as well as the energy eigenvalues are collected for all channels for convenience at \href{https://github.com/chrisculver/Pipi_Energies_Covariances}{\texttt{https://github.com/chrisculver/Pipi\_Energies\_Covariances}}.}, which as we found, make considerable contribution to the total $\chi^2$. 

Overall, the data consist of 95 energy eigenvalues in the $I=0,1,2$ channels, extracted from 6 ensembles as in Table~\ref{table:gwu_lattice} and depicted in Figure~\ref{fig:DNA}. For each of the ensembles the $m_\pi$ and $f_\pi$ have been recorded including the corresponding correlations. Varying the pion mass can lead to non-negligible effects as noted in Ref.~\cite{Bulava:2016mks}. In relation to the recorded pion mass we noted a systematic effects in $\mathcal{E}_2$, i.e., its different central value in Table~\ref{table:gwu_lattice}. Thus, we exclude this ensemble from further fits. This leaves us with 88 data points and correlations between different channels, energy levels, pion masses and decay constants within each ensemble. In exploratory fits we have found that the value of $l^r_3$ does not lead to any notable improvement of the $\chi^2$ so that we fix it to the value reported by FLAG~\cite{Aoki:2019cca}, i.e., $l^r_3=8.94\cdot10^{-6}$. This leaves us with three fit parameters ($l^r_1$, $l^r_2$, $l^r_4$) of the model as well two pairs $(m_\pi,f_\pi)_{\rm heavy/light}$ for heavy and light ensembles, respectively. We allow the pertinent four parameters to vary instead of using their central values due to the high precision of the data especially in the threshold proximity. In each fit scenario the correlated $\chi^2$,
\begin{align}
\chi^2\equiv\sum_{i=1}^{6} (X_i-Y_i)^T\cdot {\rm Cov}_{\mathcal{E}_i}^{-1}\cdot (X_i-Y_i)
\end{align}
is minimized with respect to the seven fit parameters. Here $X_i$ denotes the vector containing central values of $m_\pi$, $f_\pi$ and energy eigenvalues of a given lattice ensemble. The vector $Y_i$ contains the corresponding result obtained from the model as in Eq.~\eqref{eq:mIAM-FVspectrum}. The  covariance matrix ${\rm Cov}_{\mathcal{E}_i}$ contains uncertainties and correlations of all data, including correlations between different isospin channels.

\begin{figure*}[t!]
    \centering
    \includegraphics[width=0.32\linewidth, trim=0 0 0.95cm 0,clip]{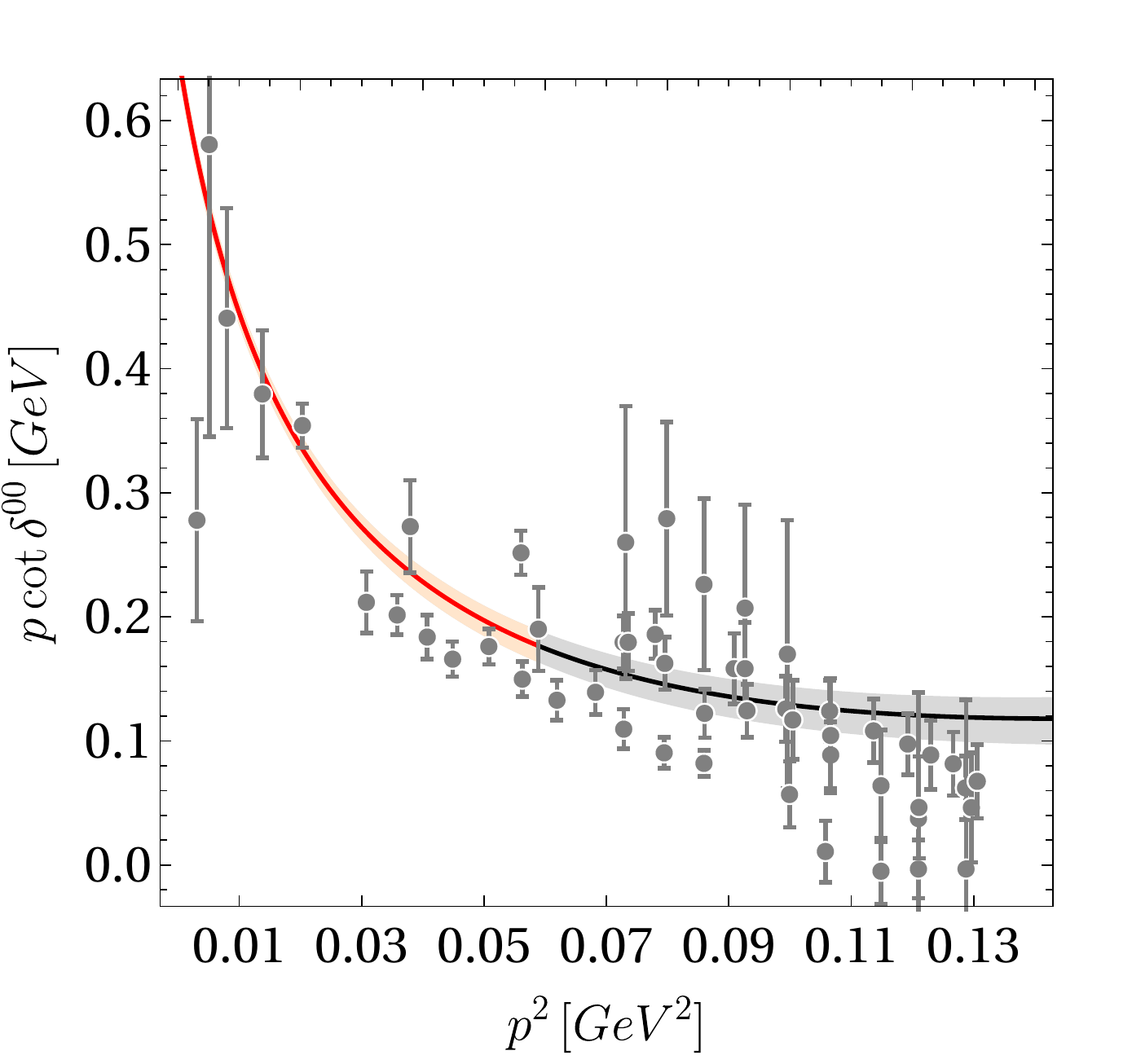}
    \includegraphics[width=0.32\linewidth, trim=0 0 0.95cm 0,clip]{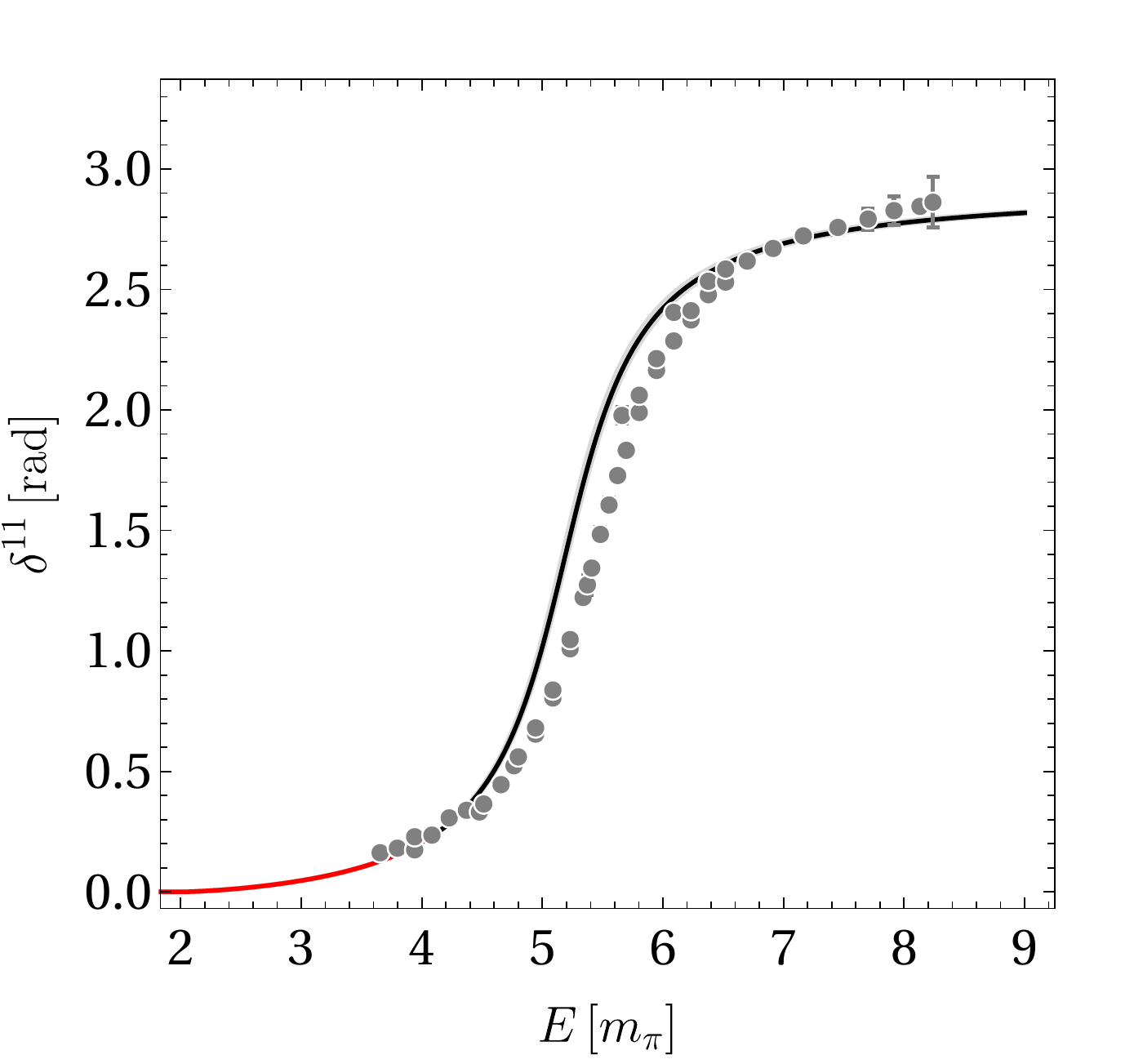}
    \includegraphics[width=0.33\linewidth, trim=0 0 0.95cm 0,clip]{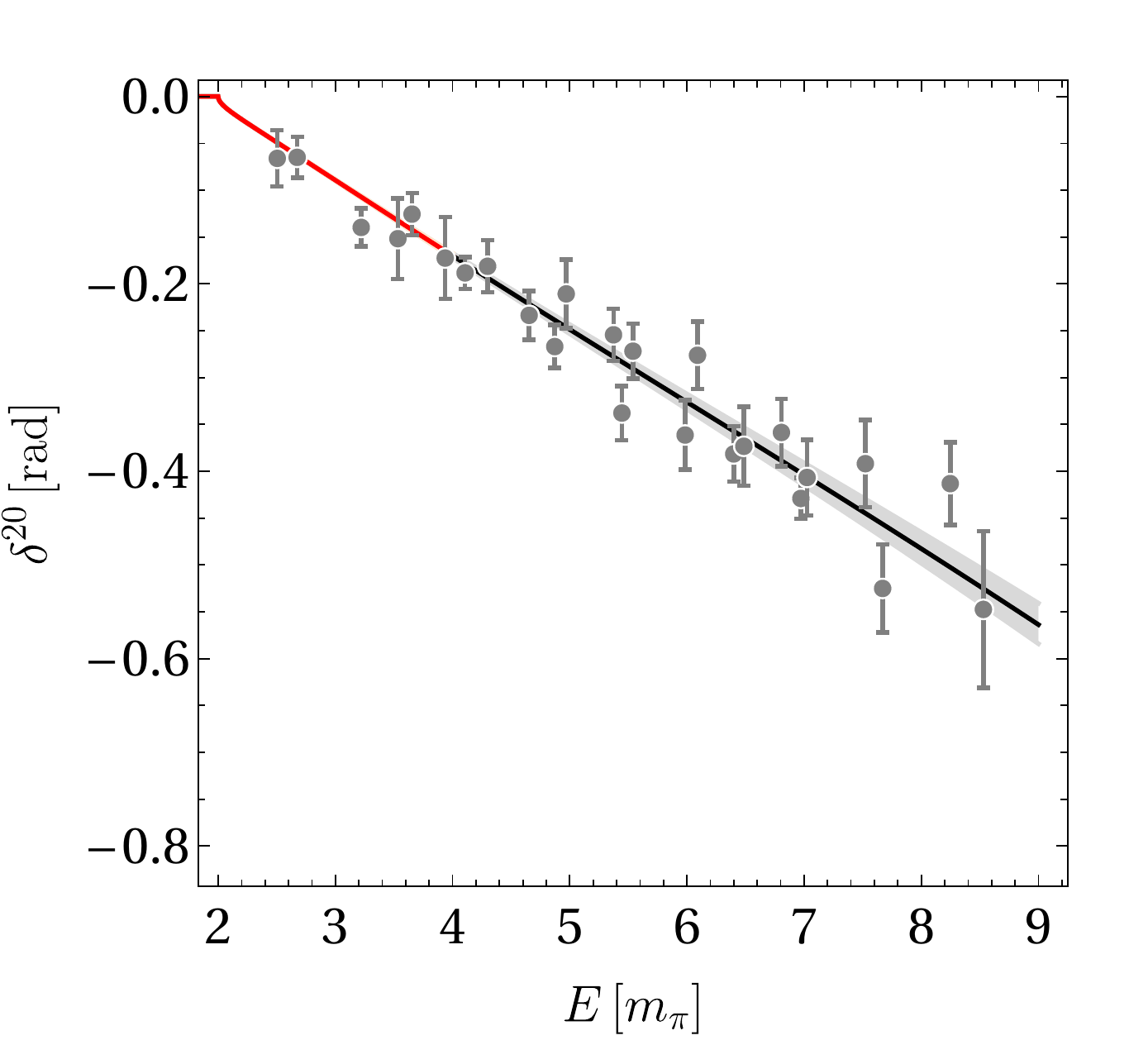}
    
    \caption{Prediction of phase-shifts in all pion-scattering channels at physical pion mass using mIAM and parameters fitted to the lattice QCD data only (Table~\ref{tab:lec}). Error bands are determined from a re-sampling routine, while the gray bars denote phase-shifts extracted from experiment
    ~\cite{Batley:2010zza,Froggatt:1977hu,Estabrooks:1974vu,Hyams:1973zf,Protopopescu:1973sh,Grayer:1974cr,Rosselet:1976pu,Janssen:1994wn,Estabrooks:1974vu} for comparison. Predictions above the inelastic threshold are grayed out.}
    \label{fig:chiralextrapolation}
\end{figure*}

\subsection{Fit results}\label{sec:fit-results}

The best fit gives $\chi^2_{\rm dof}=218/(88-7)~(\approx2.7)$  with the parameters collected in Table~\ref{tab:lec}. We explored various other fit scenarios to determine the root of the relatively large $\chi^2$ value. For instance, we note that a sizable contribution to the total $\chi^2$ is generated by the correlations within each ensemble across different isospin channels (up to 50\%). Furthermore, we have performed separate fits for light and heavy ensembles only. The best fit results of those are $\chi^2_{\rm light}=63.8/(47 - 5)\approx1.5$ and $\chi^2_{\rm heavy}=107.2/(43 - 5)\approx2.8$. The tension between the heavy and light mass fits, noted in Ref.~\cite{Culver:2019qtx} (i.e. $\chi^2_{{\rm light\&heavy}}/(\chi^2_{{\rm light}}+\chi^2_{{\rm heavy}})\approx2$), is reduced when all channels are considered. Moreover, the fact of $\chi^2_{\rm heavy}>\chi^2_{\rm light}$ is easy to understand since the model used relies on the chiral amplitudes up to the fourth chiral order. Thus, it is not too surprising that the corresponding description begins to fail at the heavier pion mass. In this respect, see the discussion in Ref.~\cite{Durr:2014oba}. Also, as depicted in Figure~\ref{fig:DNA}, large contributions to the $\chi^2$ come from energy eigenvalues at high energies in $I=1$ and $I=2$ channels, where the error bars are especially narrow. A similar observation was also made in Ref.~\cite{Guo:2016zos} dealing with the $I=1$ channel only, supported by two simpler  (Breit-Wigner and Lippmann-Schwinger type) models.

The probability distribution function of the low-energy constants is depicted in Figure~\ref{fig:ldistribution} together with that of the re-sampled $\chi^2_{\rm res}$ in the inset. The latter should follow a non-central $\chi^2$ distribution function with non-centrality parameter $\lambda=\chi^2=218$ of the fit to the original data and $k=88-7$ degrees of freedom, peaking at a re-sampled $\chi_{\rm res}^2$ of around $\chi_{\rm res}^2=300$ shown as the blue curve overlaying the histogram in the inset of Figure~\ref{fig:ldistribution}. The difference between histogram and theoretical expectation for the $\chi_{\rm res}^2$ distribution is explained by the mentioned shortcoming of the fit at higher energies / pion mass and across different isospin channels. The errors on LECs are on the order of 5-10\%, which is an order of magnitude smaller than those from the fit of the same model to one channel ($I=2$) only~\cite{Culver:2019qtx}.  This shows that inclusion of all isospin channels indeed restricts the model and thus the extraction of physically relevant information strongly.
 
Finally, the results of the fit in all three channels at two pion masses are also presented in terms of infinite-volume quantities in the lower part of Figure~\ref{fig:DNA}. There the energy eigenvalues obtained from a lattice calculation are also mapped to the phase-shifts using L\"uscher's method. In terms of the phase-shifts some of the results produce error bars wrapping through the whole codomain of this mapping. This occurs when the error on energy eigenvalue overlaps with a non-interacting value.

\begin{table*}[t!]
\centering
\begin{ruledtabular}
 \begin{tabular}{l|lll}
   \addlinespace[0.3em]
    $m_\pi$ [MeV]
   &\multicolumn{1}{l}{$\sim315$}
   &\multicolumn{1}{l}{$\sim224$}
   &\multicolumn{1}{l}{$139$}\\
   \addlinespace[0.3em]
   \hline
   \addlinespace[0.3em]
    $m_\pi\,a_0^{I=0}$
   &$+1.9008^{+0.0521}_{-0.0593}$
   &$+0.6985^{+0.0010}_{-0.0015}$
   &$+0.2132^{+0.0008}_{-0.0009}$\\
   \addlinespace[0.3em]
   \hline
   \addlinespace[0.3em]
    $m_\pi\,a_0^{I=2}$
   &$-0.1538^{+0.0021}_{-0.0018}$
   &$-0.0952^{+0.0010}_{-0.0009}$
   &$-0.0433^{+0.0002}_{-0.0002}$\\
   \addlinespace[0.3em]
   \hline
   \addlinespace[0.3em]
    $m_\sigma$ [MeV]
   &$+591^{+6}_{-5}-i109^{+4}_{-4}$
   &$+502^{+4}_{-4}-i175^{+6}_{-5}$
   &$+443^{+3}_{-3}-i221^{+6}_{-6}$\\
   \addlinespace[0.3em]
   \cline{1-1}
   \addlinespace[0.3em]
    $g_{\sigma\pi\pi}$ [MeV]
   &$533^{+2}_{-2}$
   &$426^{+2}_{-2}$
   &$397.8^{+0.6}_{-0.6}$\\
   \addlinespace[0.3em]
   \hline
   \addlinespace[0.3em]
    $m_\rho$~[MeV]
   &$+789^{+1}_{-1}-i20^{+0}_{-0}$
   &$+738^{+2}_{-1}-i43^{+1}_{-1}$
   &$+724^{+2}_{-4}-i67^{+1}_{-1}$\\
   \addlinespace[0.3em]
   \cline{1-1}
   \addlinespace[0.3em]
    $g_{\rho\pi\pi}$ [MeV]
   &$226^{+2}_{-2}$
   &$282^{+3}_{-2}$
   &$323^{+5}_{-3}$\\
   \addlinespace[0.3em]
 \end{tabular}
\end{ruledtabular}
\caption{Scattering lengths and pole positions from the global mIAM fit to lattice QCD data~\cite{Culver:2019qtx,Guo:2018zss,Guo:2018zss}. The last column shows the extrapolation to the physical point. Error bars are determined in a re-sampling routine corresponding to the parameters quoted in Table~\ref{tab:lec}. The couplings $g$ quoted in parentheses are defined as residua at the complex pole positions at $s^*=m_\sigma^2$ or $s^*=m_\rho^2$ via
$g^2=\lim_{s\to s^*}|(s-s^*)T(s)|$.}
\label{tab:phys-observables}
\end{table*}

\subsection{Predictions at the physical point}\label{sec:physpoint}

The parameters of the model $(l^r_1,l^r_2,l^r_4)$ have been determined in a fit to the lattice results in a model with correct (up to second chiral order) pion mass dependence. This allows us to extrapolate the amplitude to the physical point to confront the pertinent phase-shifts with the phenomenological results. The  extrapolations are shown in Figure~\ref{fig:chiralextrapolation} together with the phase-shifts extracted from experiment. The bands show the $1\sigma$ region, originating from the re-sampling of the  fits. To emphasize that unitarity is strictly fulfilled only up to the first inelastic threshold ($4m_\pi$), we use a different color for the predicted curves above this region. 

We observe that in the even isospin channels the prediction agrees with the experimental data very well in the elastic region and even beyond it. In the $I=1$ channel the functional behaviour is very similar to the one suggested by experiment, but is shifted to the left. This suggests a lighter mass of the $\rho$ resonance and we will return to this discussion point below.

The physical parameters, such as scattering lengths, resonance pole positions  and couplings have been discussed in length in the previous papers, dealing with single channels~\cite{Hu:2016shf,Guo:2018zss,Hu:2017wli,Hu:2016shf,Guo:2016zos}. There the dependence on the utilized model has been discussed using a broad class of chiral unitary models, Breit-Wigner type, and models based on conformal mapping. Given  that the mIAM is a better compromise between constraints from chiral symmetry~\cite{Bruns:2017gix} and analytic properties of scattering amplitudes we simplify the discussion here by discussing the results of this approach only. The collection of observables at physical and both unphysical pion masses can be found in Table~\ref{tab:phys-observables}.

The sizes of the even-isospin scattering lengths are slightly smaller than the phenomenological values $m_\pi\,a_0^{I=0}=0.2198(46)_{\rm stat}(16)_{\rm syst}(64)_{\rm th}$ and $m_\pi\,a_0^{I=2}=-0.0445(11)_{\rm stat}(4)_{\rm syst}(8)_{\rm th}$ 
of Ref.~\cite{Caprini:2011ky}. Interestingly, most lattice QCD based determinations of the latter tend to be smaller in magnitude than the results based on experimental results, see the discussion in the FLAG report~\cite{Aoki:2019cca}. 

The extrapolated pole position of the isoscalar resonance agrees well in the real part but is too small in the imaginary part when compared to the phenomenologically driven analysis of Ref.~\cite{Pelaez:2015qba}: $m_\sigma=449_{-16}^{+22}-i275_{-12}^{+12}$~MeV which is an average of analyses based on Roy equations and related methods~\cite{Colangelo:2001df, Caprini:2005zr, Moussallam:2011zg, GarciaMartin:2011jx}. For the chiral extrapolations of the HadronSpectrum data~\cite{Briceno:2016mjc} in Ref.~\cite{Doring:2016bdr} and the GW lattice QCD data~\cite{Guo:2018zss} similarly narrow $\sigma$ pole positions were found; also, in fits to experimental data only, e.g., in Refs.~\cite{Oller:2003vf, Dobado:1996ps, Doring:2011nd} based on the mIAM or unitarized ChPT with contact terms only, the $\sigma$ was rather narrow compared to the averaged value of Ref.~\cite{Pelaez:2015qba} from Roy equations. These observations suggest that methods based only on s-channel unitarization (with up to one loop in the $t,u$-channels) tend to produce slightly narrower $\sigma$ resonances compared to the Roy equations which provide better sub-threshold analytic properties. 

The extrapolated result on the isovector resonance agrees well in its width with the phenomenological value~\cite{Tanabashi:2018oca} of $\Gamma\approx150$~MeV. However, its mass is too small by $\sim40$~MeV. This corroborates a similar finding for the chiral extrapolation of several $N_f=2$ lattice QCD calculations in Ref.~\cite{Hu:2016shf} based on a simpler model and using only data from the $I=1$ channel. In this model only NLO contact interactions were used in the unitarization, based on Ref.~\cite{Oller:1998hw}. As shown here, the light $\rho$ is not a consequence of this simplification.

Another potential reason for the discrepancy between the physical $\rho$ mass and our extrapolation can be attributed to the ambiguity in determining the lattice spacing. We argue here that this issue is a bit more subtle: the lattice spacing is not directly relevant while the definition of the physical point is. To see this note that all fits are carried using inputs in lattice units: energy levels, pion masses and decay constants, and their correlations are all dimensionless. The lattice data together with the mIAM predicts the phase-shifts $\delta^{IJ}(l_i^r, E/m_\pi, f_\pi/m_\pi, \mu/m_\pi)$, all parameters being dimensionless ratios. The LEC's $l_i^r$ are fixed by the fit to the dimensionless lattice data. The dependence on $\mu/m_\pi$ is very small and can be disregarded. The prediction of $\delta^{IJ}$ versus $E/m_\pi$ is completely independent from the lattice spacing $a$. The only relevant parameter is $f_\pi/m_\pi$, which needs to be set to the value corresponding to the physical point. For the results in Fig.~\ref{fig:chiralextrapolation} we define the physical point by setting $f_\pi/m_\pi=92.4/139$, the physical values for $\pi^+$. Note that there is an ambiguity in defining this point for $N_f=2$ simulations due to the absence of the strange quark and isospin breaking effects. 

Finally, we have also checked the effects on the pion mass due to the elongation of the box. We could not detect any systematic shift of the pion mass when allowing for different fit parameters for the pion masses in the different spatial elongations. In particular, the pertinent independent-mass fit leads to a very similar chiral extrapolation.
Other possible explanations for the discrepancy could be related to the missing $K\bar K$ channel~\cite{Hu:2016shf} or higher orders effects in the chiral expansion~\cite{Niehus:2019nkl}.

\section{Summary and conclusions}\label{sec:summary}

We study pion-pion elastic scattering across all three isospin channels in two-flavor dynamical lattice QCD. It is the first time such a cross-channel analysis of lattice QCD scattering data is attempted. 
We use elongated lattices which offer a cost-effective alternative to cubic lattices in mapping out the momentum dependence in scattering processes. We consider six ensembles with elongations up to a factor of two in one of the spatial dimensions, and two quark masses corresponding to pion masses at 315 MeV and 224 MeV. The lattice input include two-pion states at rest and also states boosted along the elongated direction for enhanced momentum coverage. 

To make contact with phenomenology, we put the finite-volume spectrum across all three channels through a correlated global analysis using the inverse amplitude method. This method unitarizes chiral perturbation theory in the s-channel, matching the chiral $\pi\pi$ scattering amplitude to next-to-leading order, and it allows for  chiral extrapolation of lattice data to the physical point over a wide energy range. We treat the pion mass and decay constant as fit parameters and include their full correlations with the energy eigenvalues in the corresponding analysis; likewise, full correlations across different isospin channels are included. 

The model qualitatively captures the scattering phase-shifts in the
entire elastic region. However, the $\chi^2$  indicates
a tension between the model and the lattice QCD data. Remarkably, a large contribution to the $\chi^2$ originates from correlations between different isospin channels, which have been taken into account for the first time in the present study. Quantitatively the agreement is better for the lower quark mass and lower energies as expected from a ChPT-based model.
The cross-channel fit provides better constraints on the model parameters and this in turn leads to a tighter extrapolation to the physical point. The extrapolation agrees well with the experimental phase-shifts in the $I=0$ and $I=2$ channel. In the $I=1$
channel, the extrapolations favor a lower mass for the $\rho$ resonance, as was found in previous studies. We argued here that this
disagreement is not related to the determination of lattice spacing, since this has almost no effect on the determination of the
model parameters. Among other possible explanations for the discrepancy, there is, however, an ambiguity in defining the physical point since the experimental data is not directly comparable with $N_f=2$ simulations. 

Overall our results demonstrate that elongated lattices combined with a global analysis in the inverse amplitude method can be an effective tool in probing hadron-hadron scattering processes from first principles. 
We plan to extend the approach to more systems, 
such as three pions above threshold, and pion-baryon scattering.

\begin{acknowledgments}
We thank Dehua Guo for helping us compute the correlation between the energies in the $I=2$ channel and the $I=0,1$ channels.
CC, AA, and FXL are supported in part by the National Science Foundation CAREER grant PHY-1151648 and by U.S. DOE Grant No. DE-FG02-95ER40907. MM and MD are supported in part by the National Science Foundation CAREER grant PHY-1452055 and by U.S. DOE Grant No. DE-AC05-06OR23177. AA gratefully acknowledges the hospitality of the Physics Department at the University of Maryland where part of this work was carried out. The computations were performed on the GWU Colonial One computer cluster and the GWU IMPACT collaboration clusters. 

\end{acknowledgments}

\bibliography{ALL-REF.bib}

\end{document}